\documentclass[11pt]{article}
\usepackage{moriond,epsfig}

\def\be{\begin{equation}}
\def\ee{\end{equation}}
\def\bea{\begin{eqnarray}}
\def\eea{\end{eqnarray}}

\begin{document}
\vspace*{4cm}
\title{Measurement of Rare Kaon Decay $K^+\rightarrow\pi^+\nu\bar{\nu}$}

\author{ Shaomin Chen }

\address{TRIUMF, 4004 Wesbrook Mall, Vancouver,\\
B.C., V6T 2A3, Canada}

\maketitle\abstracts{A decade long search for the rare kaon decay
to $\pi^+\nu\bar{\nu}$ has been pursued by E787. Two signal events
are observed, giving a measurement of
the branching ratio 
Br($K^+\rightarrow\pi^+\nu\bar{\nu}$)=$1.57^{+1.75}_{-0.82}\times
10^{-10}$ and a constraint on the Cabibbo-Kobayashi-Maskawa (CKM)
matrix element $0.007<|V_{td}|<0.030$ (68\% C.L.).
}

\section{Introduction}

In the Standard Model (SM) the transition $K^+$ to $\pi^+
\nu\overline\nu$ is a Flavor Changing Neutral Current (FCNC) process in
which the first order weak 
decay is forbidden by the GIM mechanism
but is allowed, though highly suppressed, in second order due to the
differing masses of the up, charm and top quarks in the mediating
loops. Theoretically, this decay is sensitive to
the CKM matrix element $V_{td}$, which is one of two CKM matrix
elements containing CP violation information in
the SM.
A study of this decay can provide a constraint on the range of  
$V_{td}$ from the branching ratio of
$K^+\rightarrow\pi^+\nu\bar{\nu}$. The SM predicts the
branching ratio of this decay mode\cite{brk} to be
($0.72\pm0.21$)$\times 10^{-10}$ from a fit using
$B_d-\bar{B}_d$ and $B_s - \bar{B}_s$ mixing and other relevant
SM parameters. The theoretical
uncertainty in Br($K^+\rightarrow\pi^+\nu\bar{\nu}$) is small
($\sim 7$\%)~\cite{buras}, and long distance contributions to this decay
are found to be negligible~\cite{ldc}.
Therefore, a precise measurement 
 can also serve as a probe for new physics beyond
the SM.
Most interestingly, this decay together
with
the neutral kaon decay $K_L\rightarrow\pi^0\nu\bar{\nu}$ and
the third most accurately measured CKM matrix element $V_{cb}$ can also
form a unitarity triangle in the $\rho-\eta$ plane, thus providing an 
approach for understanding
CP violation.

The E787 experiment at Brookhaven National Laboratory
found the first candidate event in 1995 data~\cite{prl9597} and
recently found the second candidate event in 1998
data~\cite{prl01}. This talk will present the final E787 analysis result
using data produced from about $6\times 10^{12}$ charged kaons
at the Alternating Gradient Synchrotron (AGS) accelerator.

\section{The E787 Detector}

\begin{figure}
\begin{center}
\epsfxsize 15cm
\epsffile{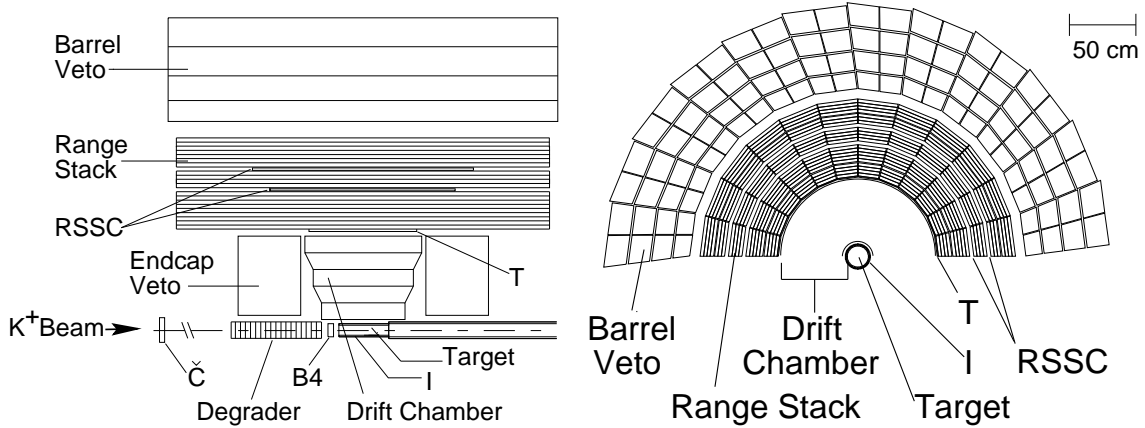}
\end{center}
\caption{Top half of side (left) and end (right) views of the E787 detector.}
\label{det}
\end{figure}

The E787 experiment (see Figure~\ref{det})
is designed to capture the signature of 
$K^+\rightarrow\pi^+\nu\bar{\nu}$, i.e., a charged kaon decay to a
charged pion of momentum below 227 MeV/$c$ and no other
associated observable product. Potential background can be
$K^+\rightarrow\mu^+\nu_\mu(\gamma)$ (due to $\mu^+$ misidentified as
$\pi^+$ or mismeasured kinematics or missed photon),
$K^+\rightarrow\pi^+\pi^0$ (due to missed photon or
 mismeasured
kinematics), beam backgrounds (due to incoming $\pi^+$
mis-identified as $K^+$ or $\pi^+$ faking $K^+$ decay at rest
or $K^+$ decay in flight or two incoming beam particles), 
or charge exchange background (due to $K^+ n\rightarrow K^0 p$,
$K^0_L\rightarrow\pi^+ l^-\nu_l$). A detailed description of the
E787 detector can be found elsewhere~\cite{e787}. The features
relevant for this analysis are outlined in the following.

When 700-800 MeV/$c$ charged kaons with about 20\% pion contamination
are delivered to the E787 detector, they pass through a set of 
beam counters. The first of these is the threshold \v Cerenkov counter
used to identify kaons and pions in the beam. The beam then passes
through two beam wire chambers (BWPC) used for identifying multiple beam
particles close to each other in space and time. The 
beam is slowed down by a degrader made of
beryllium oxide followed by lead glass, the latter
used for detecting pions in the  beam
or photons from kaon decay in the target. Located between the degrader and
the target is a hodoscope (B4 counter) consisting of
2 planes of 8 scintillator fingers each, 
which provides  $dE/dx$, position, and timing information
for the incident kaon, as well as 
K/$\pi$ separation and identification of possible two beam
particles. 

After passing through the beam counters, kaons are slowed down
and finally stopped at the center of the detector through 
ionization
energy loss in the target, which is made of 413 5-mm-square plastic
scintillating fibers, each 310 cm long and connected to a phototube.
Pulses from the phototubes are fed to ADCs, TDCs and 500-MHz CCD
transient digitizers. The target detector can also be used to
identify possible two beam particles, and to distinguish $K^+$ and $\pi^+$
using both time and energy measurements. 

When a kaon stops and decays in the target, the daughter charged
particle will pass through the I-counters, which consist of six
scintillators in a ring surrounding the target. The time measurement
from the I-counters together with the time measurement from the
beam counters are used to form a delayed coincidence requirement
in the online trigger  which rejects
prompt beam backgrounds. 

After the I-counter, the charged particle is tracked by the
Ultra Thin Chamber (UTC), which  has 12 layers of anode wires for
measuring the transverse momentum in the 1-T magnetic
field. In addition to this, there are six foils etched with helical cathode
strips providing a dip angle or $z$ measurement in the $r-z$ plane.
After  correction for energy loss in the target and
I-counter, the momentum resolution
is measured to be $\sigma_P/P\sim 1.1$\%. 

Upon exiting the UTC, the charged particle enters 
the range stack of plastic scintillators (RS), which consists of 21
radial layers in 24 azimuthal sectors. Each range stack module is
instrumented with a phototube at each end. Pulses from these
phototubes are delivered
to ADCs and 500-Mhz transient digitizers (TDs), thus providing
kinetic energy and range measurements.
Located after layer 10 and 14 are two layers of straw-tube tracking
chambers (RSSCs), which provide position measurements
of charged tracks in the RS. After making corrections for the energy
loss in the sub-detectors before entering the
RS, the range and kinetic energy resolutions are measured to be 
$\sigma_R/R\sim 2.9$\% and $\sigma_E/\sqrt{E\mbox{(GeV)}}\sim 1.0$\%,
respectively. A unique feature for the RS is
a measurement of the $\pi^+\rightarrow\mu^+\rightarrow e^+$ decay
sequence from the TDs in the range stack module in
which the $\pi^+$ comes to rest.
The decay sequence observation is a powerful
tool in identifying a charged pion. Muon rejection from this information
can reach about $10^5$. This cut is
independent of the $\pi^+/\mu^+$ separation using another
 cut on the range
and momentum correlation for different particles, where the 
$\pi^+/\mu^+$ separation is more than 3$\sigma$.

The outermost detectors are the barrel and end-cap photon vetoes. Together with
the lead glass beam counter and additional calorimeters for filling
minor openings along the beam direction, they
provide a 4$\pi$ solid angle for detecting photon activity. 

\section{Background study}

The study of $K^+\rightarrow\pi^+\nu\bar{\nu}$ adopts the technique of 
blind analysis, in which selection criteria (cuts) are designed from a study of
background samples to avoid bias. Before looking into
the signal region (``opening the box"), background levels are
estimated using the
knowledge outside the signal region. Cuts are designed in such a way
to bring the background to the 0.1 event level.  

\begin{figure}
\epsfxsize 15cm
\epsfysize 7.5cm
\begin{center}
\epsffile{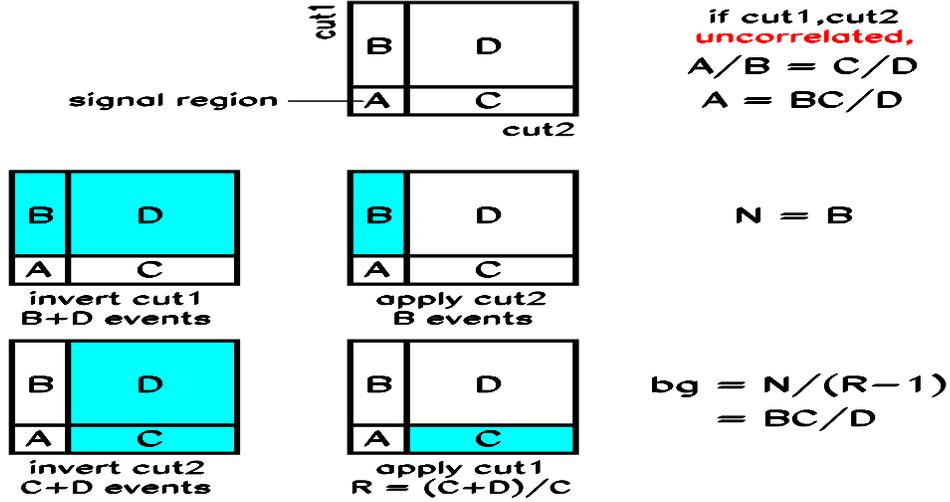}
\end{center}
\caption{The philosophy of bifurcated analysis.}
\label{bifur}
\end{figure}

To get a reliable background estimate, a so-called bifurcated analysis
is conducted. The philosophy of this method can be illustrated in
Figure~\ref{bifur}.
Experimentally, two uncorrelated cuts giving large background
rejection are selected to perform this bifurcated analysis. Based
on the event numbers in region C, B and D, the background level
in signal region A can be estimated if these two cuts are
uncorrelated. The cuts used for the bifurcated analyses are listed in
Table~\ref{cuts}. In the $\pi^+\nu\bar{\nu}$ analysis,
the $\pi^+\pi^0$ and $\mu^+\nu_\mu(\gamma)$ are the two major
backgounds. Since kinematic cuts are based the measurements of
the range, kinetic energy and momentum, which are independent of
the detection of photon activity and the particle ID using the
TD information for recognizing the
$\pi^+\rightarrow\mu^+\rightarrow e^+$ decay sequence, the bifurcated
analyses can be performed between them.
\begin{table}
\caption{Cuts used in the bifurcated analyses for the background.}
\label{cuts}
\vspace{0.4cm}
\begin{center}
\begin{tabular}{|l|l|l|}\hline
Background    & CUT1 & CUT2 \\ \hline 
$\pi^+\pi^0$ & Photon veto &Kinematic cuts\\ \hline
$\mu^+\nu_\mu(\gamma)$ & RS TD PID & Kinematic cuts\\ \hline
1 beam bkg & Target timings&B4 dE/dx cuts\\ \hline 
2 beam bkg & BWPC cuts&B4 2-hit cuts\\ \hline
\end{tabular}
\end{center}
\end{table}
To check if they are uncorrelated, a so-called
outside-the-box study is conducted by loosening the two cuts and
checking if the background level estimated 
is consistent with the observed. 

In estimating background level, $\pi^+\nu\bar{\nu}$ data are divided
into 1/3 and 2/3 samples.
The 1/3 sample is used for cuts tuning and initial
background evaluation. When the background levels are satisfactory,
all the cuts are applied to the 2/3 sample and the background
levels are re-estimated. If all cuts are set without bias, the
background level from the 1/3 and 2/3 samples should be consistent. 

When the checks on the correlation of the two cuts in the
bifurcated analysis and the consistency of background
estimates between 1/3 and 2/3 sample were performed, no
correlations or inconsistencies were observed. The final background
estimates using the 2/3 sample are given in Table~\ref{bkg}.
\begin{table}
\caption{Background estimates for 1995-97 and 1998 data.}
\label{bkg}
\vspace{0.4cm}
\begin{center}
\begin{tabular}[]{|l|l|l|}\hline
 Background   & 1995-7 & 1998\\ \hline
$\pi^+\pi^0$ & $0.0216 \pm 0.0050$& $0.0120 ^{+0.0031}_{-0.0042}$ \\ \hline
$\mu^+\nu_\mu$ &                    & $0.0092 \pm 0.0067$ \\
$\mu^+\nu_\mu\gamma$&                    & $0.0245 \pm 0.0155$ \\
$\mu^+\nu_\mu(\gamma)$
&$0.0282 \pm 0.0098$& $0.0337 ^{+0.0435}_{-0.0240}$ \\ \hline
1 beam bkg    & $0.0054 \pm 0.0042$& $0.0039 \pm 0.0012$ \\ \hline
2 beam bkg    & $0.0157 \pm 0.0149$& $0.0004 \pm 0.0001$ \\ \hline
CEX&$0.0096 \pm 0.0068$&$0.0157 ^{+0.0050}_{-0.0044}$ \\ \hline
Total  & $0.0804 \pm 0.0201$&$0.0657 ^{+0.0438}_{-0.0248}$ \\ \hline
\end{tabular}
\end{center}
\end{table}
The charge exchange background estimate is from Monte Carlo.
The kaon regeneration rate and beam profile are from the actual
measurement using data for the process of
$K^+n\rightarrow K^0_Sp, K^0_S\rightarrow\pi^+\pi^-$.

\section{Search for signal}

\begin{figure}
\begin{center}
\epsfxsize 10cm
\epsffile{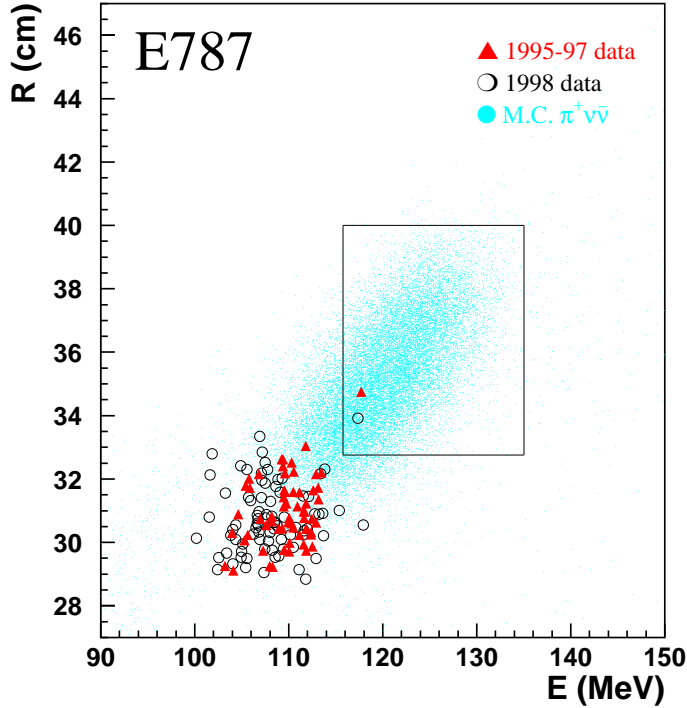}
\end{center}
\caption{Range versus kinetic energy distribution for all E787
data with all cuts applied except for the range and kinetic energy
cuts.}
\label{box}
\end{figure}

Since the background level of 0.15 event estimated for the
full E787 data is small, satisfying the single event search in
the signal region, the box is opened. 
Figure~\ref{box} shows the range versus kinetic energy for the events
surviving all the selection criteria. The box indicated by the solid
lines depicts the signal search region. Two signal events are found
inside this signal region and the events outside this box are
from the $K^+\rightarrow\pi^+\pi^0$ background due to photons escaping
 detection. Detailed studies of the candidate events  as well as a signal probability study showed that
they are consistent with the signature of
$K^+\rightarrow\pi^+\nu\bar{\nu}$.

The E787 experiment can also perform a search below the kinematic peak
of the decay~\cite{pnn2} $K^+\rightarrow\pi^+\pi^0$. The study shows
the search in this region is limited by the background from
$K^+\rightarrow\pi^+\pi^0$, which should be improved in the ongoing
E949 experiment by increasing the photon veto capability.

\section{Acceptance and branching ratio}

The acceptance is estimated using $\mu^+\nu_\mu$, $\pi^+\pi^0$, and 
$\pi^+$-scattering monitor samples taken simultaneously with the
$\pi^+\nu\bar{\nu}$ trigger and by means of a Monte Carlo $\pi^+\nu\bar{\nu}$ sample.
Table~\ref{acc} gives the acceptances for each cut category and the
final acceptance. The acceptances for $\pi^+\nu\bar\nu$ phase space,
solid angle acceptance and $\pi^+$ nucleus interaction are estimated
using Monte Carlo. Also given are the total $K^+$ triggers and the
corresponding branching ratios assuming no background.
\begin{table}
\caption{Acceptance study for $K^+\rightarrow\pi^+\nu\bar{\nu}$}
\label{acc}
\vspace{0.4cm}
\begin{center}
\begin{tabular}{|l|c|c|}\hline
Category                    & 1995-97 & 1998 \\ \hline\hline
$K^+$ stop efficiency&0.704     & 0.702 \\
$K^+$ decay after 2 ns&0.850 & 0.851 \\
$\pi^+\nu\bar\nu$ phase space&0.155     & 0.136 \\
Solid angle acceptance& 0.407     & 0.409 \\
$\pi^+$ nucl. interaction& 0.513     & 0.527 \\
Reconstruction efficiency& 0.959     & 0.969 \\
Other kinematic constraints& 0.665     & 0.554 \\
$\pi^+\rightarrow\mu^+\rightarrow e^+$ decay acc.
& 0.306     & 0.392 \\
Beam and target analysis& 0.699    & 0.706 \\
Accidental loss& 0.785     & 0.751 \\ \hline
Total acceptance&0.0021&0.0020 \\ \hline
Total $K^+$ triggers ($\times10^{12}$)&3.2 &2.7\\ \hline
Br($K^+ \rightarrow \pi^+ \nu \bar\nu$)
($\times 10^{-10}$) & $1.5^{+3.5}_{-1.2}$
                       & $1.9^{+4.4}_{-1.5}$\\ \hline
\end{tabular}
\end{center}
\end{table}
The validity is checked against the branching ratio of
$K^+\rightarrow\pi^+\pi^0$ as given below
\begin{center}
\begin{tabular}{rcl}
Br($K^+ \rightarrow \pi^+\pi^0$) & = &
             $0.208\pm0.003_{stat.}$ (1995-97)\\
             & = & $0.217\pm0.003_{stat.}$ (1998)\\
             & = & $0.212\pm0.001$ (PDG)
\end{tabular}
\end{center}
The acceptances of the $K^+ \rightarrow \pi^+ \nu
\bar\nu$ and $K^+ \rightarrow \pi^+\pi^0$ decays 
should differ only in the decay phase space. Agreement with the PDG value
is observed.

To combine searches with small statistics for both signal and
background level, a statistical analysis~\cite{junk} is performed
giving the final branching ratio measurement on
$K^+\rightarrow\pi^+\pi^0$ from E787: 
\begin{center}
Br($K^+ \rightarrow \pi^+ \nu\bar{\nu}$)=
                     $1.57^{+1.75}_{-0.82}\times 10^{-10}$
\end{center}
Assuming unitarity, $\bar{m}_t(m_t)=166\pm5$ GeV/$c^2$,
$M_W=80.41$ GeV/$c^2$ and $V_{cb}=0.041\pm0.002$, one can
derive the constraint
\begin{center}
$0.007<|V_{td}|<0.030$ (68\% C.L.),
\end{center}
without requiring any knowledge of $V_{ub}$ or $\epsilon_K$.

\section{Conclusion}

The E787 experiment has pursued a decade long search for
the rare kaon decay to $\pi^+\nu\bar{\nu}$ with two signal events
observed. The branching ratio is measured to be
$1.57^{+1.75}_{-0.82}\times 10^{-10}$. The central value is
a factor of two larger than what the Standard Model predicts, though
the large uncertainty prevents any solid conclusion on
possible new physics. It is expected that the ongoing E949 experiment,
which continues the study of the $K^+ \rightarrow \pi^+ \nu\bar{\nu}$
decay at BNL will be able to collect 5 times more statistics and provide a
critical test on the Standard Model.


\section*{References}
\begin{thebibliography}{99}
\bibitem{brk}Giancarlo D'Ambrosio and Gino Isidori, Phys.Lett. B530 (2002) 108.
\bibitem{buras}A.J. Buras and R. Fleischer, hep-ph/9704376.
\bibitem{ldc}Gino Segr\`e, Phys.Rev. D61 (2000) 077301;
S. Fajfer, Nuovo Cim. A110 (1997) 397;
C.Q. Geng $et~al.$, Phys. Rev. D54 (1996) 877;
M. Lu and M.B. Wise, Phys. Lett. B324 (1994) 461.
\bibitem{prl9597}S. Adler $et~al.$, Phys. Rev. Lett. 84, (2000) 3768;
S. Adler $et~al.$, Phys. Rev. Lett. 79, (1997) 2204.
\bibitem{prl01}S. Adler $et~al.$, Phys. Rev. Lett. 88, (2001) 041803.
\bibitem{e787} M.S. Atiya $et~al.$, Nucl. Inst. and Meth. A321
(1992) 129; E. W. Blackmore $et~al.$, Nucl. Inst. and Meth. A404
(1998) 295; I.-H. Chiang $et~al.$, IEE Trans. Nucl. Sci. NS-42
(1995) 394; T.K. Komatsubara $et~al.$, Nucl. Inst. and Meth. A404
(1998) 315; M. Kobayashi $et~al.$, Nucl. Inst. and Meth. A337 (1994) 335;
D.A. Bryman $et~al.$, Nucl. Inst. and Meth. A396 (1997) 394;
M. Burke $et~al.$, IEEE Trans. Nucl. Sci. NS-41 (1994) 131;
C. Witzig and S. Adler, Real-Time Comput. Appl. (1993) 123;
S. Adler, Intl. Conf. Electr. Part. Phys. (1997) 133;
C. Zein $et~al.$, Real-Time Comput. Appl. (1993) 103.
\bibitem{pnn2} S. Adler $et~al.$, hep-ex/0201037.
\bibitem{junk}T. Junk, NIM A434 (1999) 435.
\end{thebibliography}
\end{document}